\documentclass{ws-p9-75x6-50}

\begin{document}

\title{SLOTT-AGAPE Data Processing}
\author{G. Iovane}

\address{Dipartimento di Scienze Fisiche "E.R.Caianiello", Universit\'a di Salerno, Italy
\\E-mail: geriov@sa.infn.it}
\maketitle

\abstracts{ MEDEA Program ({\bf M}icrolensing {\bf E}xperiment
{\bf D}ata-Analysis Software for {\bf E}vent with {\bf
A}mplification) is here presented. It is developed to the pixel
lensing analysis used by SLOTT-AGAPE ({\bf S}ystematic {\bf
L}ensing {\bf O}bservation at {\bf T}oppo {\bf T}elescope - {\bf
A}ndromeda {\bf G}ravitational {\bf A}mplification {\bf P}ixel
{\bf E}xperiment) Collaboration. All software components are fully
automated and they can be used on-line and off-line.}

\section{Gravitational lensing and pixel lensing technique}

In the recent surveys [1] monitoring millions of stars much attention has
been focused on the possibility that a relevant fraction of galactic dark
matter could consist of MACHOs (Massive Astrophysical Compact Halo Object).
In order to detect them, it was proposed by Paczynski (1986) to search for
dark objects by gravitational lensing [2]. In fact, when a compact object
passes near the line of sight of a background star, the luminosity of this
star increases giving rise to a characteristic luminosity curve.

In a dense field, many stars contribute to any pixel of the CCD camera at
the focal point of the telescope. When an unresolved star is sufficiently
magnified, the increase of the light flux can be measured on the pixel.
Therefore, instead of monitoring individual stars, we follow the luminous
intensity of the pixels. Then all stars in the field, and not the only few
resolved ones, are candidates for a microlensing event. Of course, only the
brightest stars will be amplified enough to become detectable above the
fluctuations of the background, unless the amplification is very high and
this occurs very seldom. In a galaxy like M31, however, this is compensated
by the very high density of stars.

If $\phi $ is the amplified flux detected by the pixel and $\phi _{B}$ the
one of the background, the flux variation is given by

\begin{equation}
\phi _{star}(A-1)=\phi -\phi _{B}
\end{equation}

and

\begin{equation}
\phi _{B}=\phi _{star}+\phi _{others}
\end{equation}

where $\phi _{star}$ is the star flux at rest (i.e., when it is
not lensed) and $\phi _{others}$ is the flux of the stars in the
same pixel. $A$ is the amplification

\begin{equation}
A=\frac{x^{2}+2}{x\sqrt{x^{2}+4}}\qquad with \qquad x^{2}=\omega
^{2}(t-t_{0})^{2}+x_{\min }^{2},
\end{equation}

where $\omega =t_{E}^{-1}$ is the inverse Einstein time; $x_{\min }$ is the
impact parameter in units of the Einstein radius $R_{E},$ and $t_{0}$ is the
time of maximum amplification.
\newpage
\section{\protect\bigskip MEDEA Software}

\subsection{Conceptual Design for data acquisition, selection and analysis}

A two level trigger follows the data acquisition to perform the
on-line selection of interesting events that can be studied
quasi-on-line. Other events, which should be more complicate
lensing events, variable stars, novae and supernovae and so on,
are not lost during the selection, but are stored for an off-line
analysis.

More in particular thanks to the first trigger level light curves
with luminosity variation are selected; while the second trigger
level is responsible of lensing event selections [3].

\subsection{MEDEA Components}

The software system is made by different units:

\begin{itemize}
\item  The Intelligent Data Acquisition \textbf{(I-DAQ)} Unit, that is
responsible of the data acquisition and technical corrections.

\item  The Control Unit, which thanks to Telescope Control System (\textbf{%
TCS}), controls the telescope by following the instruction from the DataBase
Control System (\textbf{DBCS}) or console.

\item  The DataBase (\textbf{DB}) Unit is the intelligent part of the
system; in fact, here we have the data storage and processing. Such a unit
is the natural link from the observations to the data analysis components.

\item  The Processing and Analyzing (\textbf{P\&A}) Unit is the platform
where the massive data analysis is performed. It is made of three
main sub-units:

\begin{itemize}
\item  Data pre-processing sub-unit, (\textbf{P\&A}).(\textbf{DAPP}), used
for geometrical alignment, photometric calibration, and seeing correction;

\item  Data processing sub-unit, (\textbf{P\&A}).(\textbf{DAP}),  is the
first trigger level, that selects amplified light curves by using a specific
peak detection algorithm;

\item  Data analysis sub-unit, (\textbf{P\&A}).(\textbf{DAU}), used for
testing different lensing model (single point-like lens and
source, double source, extended source), for performing color
correlation test, and for verifying the statistical
Kolmogorov-Smirnov test and Durbin-Watson one.

\end{itemize}
\end{itemize}

\end{document}